\documentstyle{article}

\def\be{\begin{eqnarray}}
\def\ee{\end{eqnarray}}
\def\nn{\nonumber}

\hoffset= - 1.0in         % switch off for draft style
\begin{document}

\hfill ITEP/TH-72/07

\bigskip

\centerline{\Large{Hamiltonian Formalism in the
Presence of Higher Derivatives}}

\bigskip

\centerline{\it A.Morozov}

\bigskip

\centerline{ITEP, Moscow, Russia}

\bigskip

\centerline{ABSTRACT}

\bigskip

A short review of basic formulas from Hamiltonian formalism
in classical mechanics in the case when Lagrangian contains
$N$ time-derivatives of $n$ coordinate variables.
For non-local models $N=\infty$.

\bigskip

\bigskip

\section{Introduction}

Theories with higher derivatives were long at the
periphery of theoretical physics.
The basic reason for this is that they do not appear in
everyday physical problems.
This in turn can have a reason:
if, as many believe these days,
observable world is described by a low-energy limit of
some yet-unknown fundamental theory,
then it is naturally governed by Lagrangian dynamics
with lowest possible -- this means a pair of first --
time-derivatives and Newton law includes only acceleration.
From this point of view there is no restriction on the
number of derivatives in the {\it fundamental} theory,
and higher derivative terms are indeed present in most approaches,
from QFT formulations of string and M-theory \cite{SFT}
to pure QFT models like asymptotically safe gravity \cite{ASG}
or (the quantum version of)
the recent $E_8$ unification model \cite{Lisi}.
It goes without saying that various non-local and
innumerable non-commutative models all fit into category
of higher-derivative theories.
More than that, even the ordinary physical theories,
like {\it classical} electrodynamics, appear inconsistent
without higher derivatives, if one includes radiation
phenomena and allows space-time dimension to be greater
than $4$ -- like one does, for example, in amusing
TeV-gravity models \cite{TVgra}.
In these circumstances the resolution of radiation friction
and electromagnetic mass "problems" requires inclusion into
the bare ("fundamental") action of counter-terms which not
only renormalize mass (as in $d=4$), but necessarily
include higher derivatives \cite{radfri}.
Of course, higher derivatives are also used for purposes
of UV regularization in more formal context,
especially in gauge invariant and supersymmetric models
\cite{HDReg},
even if inclusion of such terms is not physically unavoidable.
Last but not the least, higher-derivative terms are the
common place in all effective theories, from solid state
physics to quantum gravity.

For all these reasons the higher-derivative
dynamics is slowly gaining new attention,
see \cite{hdf}-\cite{hdl} for the relatively recent discussions
from various viewpoints,
as well as \cite{hdff}-\cite{NaHa}
for some classical papers and monographs.
It should be emphasized that this almost-untouched ground
is very attractive from the point of view of "theoretical theory"
and is intimately related to modern topological theory
\cite{ALG},
$L_\infty^{(n)}$ structures {\it a la} \cite{Lninf},
non-linear algebra \cite{nolal} etc.
Since \cite{FGM} it is known that when such theories are
required to be reparametrization invariant
(what is the case in most of thinkable applications)
new phenomena of outstanding beauty occur.
Of special interest is symplectic geometry behind such theories
\cite{DoSto,DBS}.

This short note is devoted to the 0-th chapter of
higher-derivative theory.
It contains a short list of elementary formulas -- well-known to
a narrow class of interested people ever since \cite{hdff} --
about {\it classical} Lagrangian and Hamiltonian dynamics,
which can be used for comparison with results of various
more-advanced approaches.

\section{Lagrangian formalism}

Consider the classical mechanics with the action
\be
S\{q^\alpha(t)\} = \int L dt,
\label{action}
\ee
where Lagrangian
$\ L\Big(q^\alpha, d_tq^\alpha, \ldots, (d_t^N q^\alpha)\Big)\ $
depends on the first $N$ time-derivatives
$q_i^\alpha \equiv d_t^{i}q^\alpha$ of $n$ coordinate variables
$q^\alpha = q_0^\alpha$, $\alpha = 1,\ldots,n$.
Obviously,
\be
\dot q_i^\alpha = q_{i+1}^\alpha
\label{qder}
\ee
Introduce the variational derivatives w.r.t. $q_i^\alpha$
for all $i\geq 0$:
$$\delta^{i-1}_\alpha \equiv \partial^i_\alpha -
d_t\partial^{i+1}_\alpha + d_t^2\partial^{i+2}_\alpha - \ldots,$$
where $\partial^i_\alpha = \partial/\partial q_i^\alpha$ and
the momenta
\be
\Pi^i_\alpha = \delta^i_\alpha\! L
\ee
These operators are related by time-derivatives:
\be
\dot \delta^{i-1}_\alpha \equiv
d_t \delta^{i-1}_\alpha = \partial^i_\alpha- \delta^i_\alpha
\ee
in a way, dual to (\ref{qder}).

The Euler-Lagrange equations of motion are
$$\Pi^{-1}_\alpha \equiv \delta^{-1}_\alpha L = 0.$$
These are equations of order $2N$ in time-derivatives.
Initial conditions are imposed on
$q^\alpha, d_tq^\alpha, \ldots, (d_t^{2N-1} q^\alpha)$.

Note that superscripts are {\it not} powers.
In what follows we suppress indices $\alpha$ in some
formulas.

\section{Hamiltonian formalism}

The phase space is $2Nn$-dimensional with
canonically conjugate coordinates $q_i^\alpha$ and $\Pi^i_\alpha$,
where $i=0,\ldots,N-1$. The $q_N^\alpha$ are functions
of these independent variables.

The closed symplectic 2-form
\be
\Omega = d\theta =
\sum_{i=0}^{N-1} d\Pi^i_\alpha \wedge dq_i^\alpha,
\ee
is conserved:
\be
\dot\Omega = d_t\Omega = 0,
\ee
while the time-derivative of the associated pre-symplectic 1-form
\be
\theta = \sum_{i=0}^{N-1} \Pi^i_\alpha dq_i^\alpha
\ee
is exact:
\be
\dot\theta = dL.
\ee
Hamilton equations state that
\be
\dot q_i^\alpha = q_{i+1}^\alpha =
\frac{\partial H}{\partial \Pi^i_\alpha}, \nn \\
\dot\Pi^i_\alpha = -\Pi_\alpha^{i-1} - \frac{\partial L}
{\partial q_i^\alpha} =
-\frac{\partial H}{\partial q_i^\alpha}
\label{HE}
\ee
with the Hamiltonian
\be
H(\Pi,q) = \sum_{i=0}^{N-1} \Pi^i_\alpha q_{i+1}^\alpha - L
\label{Ham}
\ee
and $q_N^\alpha$ expressed through canonical variables.

Consider
$\bar S(\bar q, \bar{\bar q})$ --
the action (\ref{action}), evaluated on the
classical trajectory with the boundary conditions
$\bar q^\alpha_i = q^\alpha_i(\bar t)$ and
$\bar{\bar q}^\alpha_i = q^\alpha_i(\bar{\bar t})$,
$i = 0,\ldots, N-1$.
Then
\be
\frac{\partial \bar S}{\partial \bar q_i^\alpha} =
-\bar \Pi^i_\alpha, \nn\\
\frac{\partial \bar S}{\partial \bar{\bar q}_i^\alpha} =
\bar{\bar \Pi}^i_\alpha
\label{varbou}
\ee

All these relations are obvious generalizations of
those in the simplest case of $N=1$, see \cite{LMI}.

\section{Proofs}

The proofs of above relations are straightforward:
\be
d_t\left[\sum_{i=0} dq_i\wedge d\delta^i\right] =
\sum_{i=0} \Big(dq_{i+1}\wedge d\delta^i - dq_i\wedge
d\delta^{i-1} + dq_i\wedge d\partial^i\Big) = \nn \\ =
-dq_0\wedge d\delta_{-1} + \sum_{i=0} dq_i\wedge d\partial^i
\ee
When acting on $L$, the first term vanishes on
equations of motion, $\delta^{-1}L = 0$, while the
second term becomes
\be
\sum_{i=0} dq_i\wedge d\Big(\partial_i L\Big) =
\sum_{i,j=0}  \frac{\partial^2L}{\partial q_i\partial q_j}
dq_i\wedge dq_j = 0.
\ee
Similarly
\be
d_t\left[\sum_{i=0} dq_i \delta^i\right] =
\sum_{i=0} \Big(dq_{i+1} \delta^i - dq_i \delta^{i-1} +
dq_i \partial^i\Big) = -dq_0 \delta^{-1} + d
\ee
Again, when acting on $L$  the first term vanishes
on equations on motion.

Hamiltonian derivatives in (\ref{HE}) are:
\be
\frac{\partial H}{\partial \Pi^i} = q_{i+1} +
\left(\Pi^{N-1} - \frac{\partial L}{\partial q_N}\right)
\frac{\partial q_N}{\partial \Pi^i}, \nn \\
-\frac{\partial H}{\partial q_i} =
-\Pi^{i-1}(1-\delta_{i,0}) +
\frac{\partial L}{\partial q_i} -
\left(\Pi^{N-1} - \frac{\partial L}{\partial q_N}\right)
\frac{\partial q_N}{\partial q_i}
\ee
The terms in brackets at the r.h.s. vanish because
$\Pi^{N-1} = \delta^{N-1}L = \partial^N L$.

Finally, the variation of action $S\{q(t)\}$ under the variation
$\delta q(t)$ of its argument is equal to
\be
\delta S = \int \sum_{i=0}^N \delta x_i \partial^i L =
\oint \delta x_j \delta^{j}L  + \int \delta^{-1}L
\ee
On classical trajectory the second term vanishes, while
the boundary contributions in the first term gives rise to
(\ref{varbou}).

\section{Towards cohomological formulation}

The key role in above calculations is played by the operator
\be
\hat A = \sum_{i=1} (-)^i\partial^{i-1}\otimes \partial^{-i} =
\frac{1}{1\otimes 1 + \partial\otimes \partial^{-1}}\,
1\otimes \partial^{-1} =
\frac{1}{\partial^{-1}\otimes\partial + 1\otimes 1}
\,\partial^{-1}\otimes 1
\ee
It is a formal inverse of $\partial$, which acts on the product
by Leibnitz rule:
$$\Big(\partial\otimes 1 + 1\otimes\partial\Big) \hat A =
\frac{1}{1\otimes 1 + \partial\otimes \partial^{-1}}
\Big(\partial\otimes \partial^{-1} + 1\otimes 1\Big)
= 1\otimes 1$$
Conceptually, for $\partial = d_t$
$$\Omega = \hat A_* dq
\begin{array}{c}\wedge\\ \otimes\end{array}
\frac{\delta L}{\delta q}.$$
Indeed, for
$$\frac{\delta}{\delta q_{i}} = \frac{\partial}{\partial q_{i}}-
\partial\frac{\partial}{\partial q_{i+1}} + \ldots$$
we have
$$\partial \delta_{i+1} = \partial_i - \delta_i$$
or
$$\frac{\delta}{\delta q_{i+1}} =
-\partial_*^{-1}\frac{\delta}{\delta q_i} = \ldots
= (-)^i\partial_*^{-i}\frac{\delta}{\delta q}$$
where $\partial_*^{-1}$ properly takes care of the
$\partial/\partial q_i$ which lies in "cohomology" of $\partial$.

A more careful treatment should take into account the
difference between $\hat A_*$ and $\hat A$.
It is this difference that
makes the above $\Omega$ non-vanishing, despite $\delta L = 0$.
Only the application of time-derivative $\partial$
eliminates $\hat A_*$, but without $\partial$ there is no
vanishing.

\section{Non-local examples}

Hamiltonian formalism is immediately applicable to arbitrary
functionals, including non-local.

For $q$-quadratic examples one can take
\be
S\{q(t)\} = \int q\frac{1}{1+M^{-2}\partial^2}q \ dt
\ee
or
\be
S\{q(t)\} = \int q(t)q(t+\epsilon) dt =
\int q(t) e^{\epsilon \partial} q(t) dt
\ee
(solutions of Euler-Lagrange equations in this case
are antiperiodic functions $q(t+2\epsilon) = -q(t)$),\\
or, in general,
\be
S\{q(t)\} = \frac{1}{2}\int
\left(\sum_{n} a_n \Big(\partial^n q\Big)^2
\right)dt
\ee
with time-independent $a_n$ (not a necessary restriction,
of course).

Then
\be
\Pi_i = \delta^{i} L = \sum_{j=0} (-)^{j}a_{i+j+1}
\partial^{i+2j+1}q
\ee
and
\be
\Omega = \sum_{i=0} dq_i\wedge d\Pi_i =
\sum_{i,j\geq 0} (-)^j a_{i+j+1} dq_i \wedge dq_{i+2j+1}
\ee
Nota that terms $dq_i\wedge dq_{i+2j}$ do not appear in
this expansion.
Since coefficients $A_{ij}$ in the matrix
\be
\Omega = \sum_{i<j} A_{ij}dq_i\wedge dq_j
\ee
are time independent, they are forced to be of the form
$A_{ij}=(-)^j A_{i+j}$
(familiar from the theory of Toda chain $\tau$-functions,
see \cite{UFN3})
by the conservation condition:
\be
\dot A_{ij} + A_{i-1,j} + A_{i,j-2} = 0
\ee

\section{Reparametrization-invariant actions}

Transformations $t\rightarrow u(t) = t + \epsilon(t)$:
\be
q^\alpha_1 \equiv \dot q^\alpha \rightarrow u q_1^\alpha
\rightarrow q_1^\alpha + \epsilon q^\alpha_1, \nn \\
q^\alpha_2 \equiv
\ddot q^\alpha \rightarrow u^2 q_2^\alpha + u\dot u q^\alpha_1 \rightarrow
q_2^\alpha + 2\epsilon q_2^\alpha + \dot\epsilon q_1^\alpha, \nn \\
%\dddot
q_3^\alpha \rightarrow u^3 q_3^\alpha + 3u^2\dot u q_2^\alpha +
u\dot u^2 q_1^\alpha + u^2\ddot u q_1^\alpha \rightarrow
q_3^\alpha + 3\epsilon q_3^\alpha + 3\dot\epsilon q_2^\alpha +
\ddot\epsilon q_1^\alpha,\nn \\
\ldots
\ee
or
\be
q_k^\alpha \equiv \partial^k_t q^\alpha \rightarrow q_k^\alpha +
\sum_{l=0}^{k-1} C_k^{k-l-1} q_{k-l}^\alpha \partial_t^{l}\epsilon
\ee

Invariance of the action means that for any $\epsilon(t)$
\be
\epsilon L =
\epsilon\left\{ \dot q^\alpha\frac{\partial}{\partial \dot q^\alpha} +
2\ddot q^\alpha\frac{\partial}{\partial\ddot q^\alpha} +
3q_3^\alpha\frac{\partial}{\partial q_3^\alpha} +
4q_4^\alpha\frac{\partial}{\partial q_4^\alpha} + \ldots \right\}L +
\nn \\
+ \dot\epsilon\left\{
 \dot q^\alpha\frac{\partial}{\partial \ddot q^\alpha} +
3\ddot q^\alpha\frac{\partial}{\partial q_3^\alpha} +
6q_3^\alpha\frac{\partial}{\partial q_4^\alpha} +\ldots \right\}L +
\nn \\
+ \ddot\epsilon\left\{
 \dot q^\alpha\frac{\partial}{\partial  q_3^\alpha} +
4\ddot q^\alpha\frac{\partial}{\partial q_4^\alpha}  +\ldots \right\}L +
\nn \\
+ (\partial^3_t\epsilon)\left\{
 \dot q^\alpha\frac{\partial}{\partial  q_4^\alpha} +\ldots \right\}L +
\ldots
\label{repinvL}
\ee
or
\be
\sum_{m=0}^\infty  K_m \partial_t^m\epsilon = 0 \ \
\ \ {\rm i.e.} \ \ \ \
%\sum_{m=0}^\infty \partial_t^m\epsilon\left\{
K_m = \sum_{k\geq 1}^\infty C_{m+k}^{k-1}  q_k^\alpha \frac{\partial L}
{\partial q_{k+m}^\alpha} - L\delta_{m,0} = 0%\right\} = 0
\ee
where binomial coefficients
$C_{m+k}^{k-1} = \frac{(m+k)!}{(k-1)!(m+1)!}$. \\

As a corollary, the ordinary Hamiltonian (\ref{Ham}) vanishes
identically:
$$
H = \sum_{i=0}^{N-1} \Pi^i_\alpha q_{i+1}^\alpha - L =
\sum_{k\geq 1} q^\alpha_k\left(\frac{\partial L}
{\partial q_k^\alpha} -
\frac{d}{dt}\frac{\partial L}{\partial q_{k+1}^\alpha} +
\frac{d^2}{dt^2}\frac{\partial}{\partial q_{k+2}^\alpha}
- \ldots\right)
- L =
$$ $$
= K_0 - \sum_{k\geq 1}\left(
(k-1)q^\alpha_k\frac{\partial L}{\partial q_k^\alpha} + %\sum_l
q^\alpha_k \frac{\partial^2}
{\partial q^\alpha_{k+1}\partial q^\beta_l}\
q^\beta_{l+1} - \ldots\right) =
$$ $$
= K_0 - \frac{dK_1}{dt} + \sum_k \left(
\frac{(k-1)(k-2)}{2}q^\alpha_k\frac{\partial L}
{\partial q_k^\alpha} +
\frac{(k-1)(k+2)}{2}q^\alpha_k
\frac{\partial^2 L}{\partial q^\alpha_{k+1}\partial q^\beta_l}
\ q^\beta_{l+1}
+ \right.
$$ $$
\left. + q^\alpha_k
\frac{\partial^2 L}{\partial q^\alpha_{k+2}\partial q^\beta_l}
\ q^\beta_{l+2}
+ q^\alpha_k \frac{\partial^3 L}
{\partial q^\alpha_{k+2}\partial q^\beta_l \partial q^\gamma_m}\
q^\beta_{l+1}q^\gamma_{m+1} - \ldots \right) =
$$
\be
= K_0 - \frac{dK_1}{dt} + \frac{d^2K_2}{dt^2} - \frac{d^3K_3}{dt^3}
+ \ldots = 0
\ee
As in every gauge invariant theory the Hamilton equations
involve the constraint $\Phi$ -- generator of gauge transformation
\be
\dot q^\alpha = \frac{\partial \Phi}{\partial\Pi_\alpha},\nn \\
\dot\Pi_\alpha = -\frac{\partial \Phi}{\partial q^\alpha}
\ee
-- instead of the naive Hamiltonian \cite{Tyu}.

A special case with no dependence on $q^\alpha_0$ and $N=1$
was  studied in \cite{FGM}.

\section{Example of $n=2$ and $N=2$}

According to (\ref{repinvL})
in this case $L$ is a function of $q_1$, $q_2$, $v_1=\dot q_1$,
$v_2 = \dot q_2$ and $z = v_1a_2-v_2a_1 = v_1\dot v_2 - v_2\dot v_1$
of definite homogeneity degree:
\be
3z\left.\frac{\partial L}{\partial z}\right|_{v_{1,2}} +
v_1\left.\frac{\partial L}{\partial v_1}\right|_{v_2,z}
+ v_2\left.\frac{\partial L}{\partial v_2}\right|_{v_1,z} = L
\ee
The momenta are equal to:
\be
\Pi_1^0 = \left.\frac{\partial L}{\partial v_1}\right|_{v_2,a_{1,2}}-
\frac{d}{dt}\left.\frac{\partial L}{\partial a_1}\right|_{v_{1,2},a_{2}}
= \left.\frac{\partial L}{\partial v_1}\right|_{v_2,z}
+a_2\left.\frac{\partial L}{\partial z}\right|_{v_{1,2}}
+\frac{d}{dt}\left(v_2
\left.\frac{\partial L}{\partial z}\right|_{v_{1,2}}\right) = \nn \\ =
\left.\frac{\partial L}{\partial v_1}\right|_{v_2,z}
+2a_2\left.\frac{\partial L}{\partial z}\right|_{v_{1,2}}
+v_2\frac{d}{dt}
\left.\frac{\partial L}{\partial z}\right|_{v_{1,2}} =
\frac{\partial L}{\partial v_1} + 2a_2\frac{\partial L}{\partial z}
+v_2\left(v_1\frac{\partial^2 L}{\partial z\partial q_1} +
a_1\frac{\partial^2 L}{\partial z\partial v_1} +
(v_1w_2-v_2w_1)\frac{\partial^2 L}{\partial z^2}\right), \nn
\ee
\be
\Pi_2^0 = \left.\frac{\partial L}{\partial v_2}
\right|_{v_1,a_{1,2}}-
\frac{d}{dt}\left.\frac{\partial L}{\partial a_2}
\right|_{v_{1,2},a_{1}},  \\ \nn
\Pi_1^1 = \left.\frac{\partial L}{\partial a_1}
\right|_{v_{1,2},a_{2}}, \ \ \ \
\Pi_1^2 = \left.\frac{\partial L}{\partial a_2}
\right|_{v_{1,2},a_{1}}
\ee

Reparametrization invariance is always, not only in this
example, represented as homogeneity
condition for a function, which depends on
peculiar combinations
\be
z_{ij} = {\dot q_i \ddot q_j - \ddot q_i \dot q_j}
\sim \partial_t\left(\frac{\dot q_i}{\dot q_j}\right), \ \ \ \ \
%\ee
%and their higher-order analogues,
%\be
z_{ij;kl} \sim \partial_t\left(\frac{\dot z_{ij}}{\dot z_{kl}}\right),
\ \ \ \ \ \
z_{ij;kl|i'j';k'l'} \sim
\partial_t\left(\frac{\dot z_{ij;kl}}{\dot z_{i'j';k'l'}}\right),
\ \ldots
\ee
which are "elementary monomials", transforming homogeneously
under the time-reparametrizations.
Of course, they
play essential role in the theory, see, for example, \cite{DBS}.

\section*{Acknowledgements}

I am indebted for numerous discussion to Valery Dolotin and
to other participants of his seminars at ITEP.

I acknowledge the hospitality of OCU and support of JSPS
during completion of this text.
This work is partly supported by
Russian Federal Nuclear Energy Agency,
by the joint grant 06-01-92059-CE,  by NWO project 047.011.2004.026,
by INTAS grant 05-1000008-7865, by ANR-05-BLAN-0029-01 project and
by the Russian President's Grant of Support for the Scientific
Schools NSh-8004.2006.2 and by RFBR grant
07-02-00645.

\end{document}